# Construction of Community Web Directories based on Web usage Data

Ramancha Sandhyarani[1], Bodakuntla Rajkumar[2] and Jayadev Gyani[3]

[1]Department of Computer Science Engineering, JNTUH, Jayamukhi Institute of Technological Sciences Narsampet, Warangal, Andhrapradesh-506332, India
`sandhya_guptha_r@yahoo.co.in`
[2]Department of Information Technology, JNTUH, Jayamukhi Institute of Technological Sciences Narsampet, Warangal, Andhrapradesh-506332, India
`raj.bodak@gmail.com`
[3]Department of Computer Science Engineering, JNTUH, Jayamukhi Institute of Technological Sciences Narsampet, Warangal, Andhrapradesh-506332, India
`jayadevgyani@yahoo.com`

## *Abstract*

*This paper support the concept of a community Web directory, as a Web directory that is constructed according to the needs and interests of particular user communities. Furthermore, it presents the complete method for the construction of such directories by using web usage data. User community models take the form of thematic hierarchies and are constructed by employing clustering approach. We applied our methodology to the ODP directory and also to an artificial Web directory, which was generated by clustering Web pages that appear in the access log of an Internet Service Provider. For the discovery of the community models, we introduced a new criterion that combines a priori thematic informativeness of the Web directory categories with the level of interest observed in the usage data. In this context, we introduced and evaluated new clustering method. We have tested the methodology using access log files which are collected from the proxy servers of an Internet Service Provider and provided results that indicates the usability of the community Web directories. The proposed clustering methodology is evaluated both on a specialized artificial and a community Web directory, indicating its value to the user of the web.*

## 1. INTRODUCTION

The objective of the paper is building the Web directories according to the preferences of user communities. A Web directory, such as Microsoft (www.microsoft.com) and the Open Directory Project (ODP) (dmoz.org), users can find any topic that they are interested by searching the websites, starting with large categories and gradually moving down and choosing the category related to their interests. Searching for a particular topic ,user has to check deep inside the directory. Hence, the size and the complexity of the Web directory has increased and information overload problem will occur. To overcome the difficulties of web directories, we focus on the construction of community web directories. The users of a community can use the community directory as a starting point for navigating the Web, based on the topics that they are interested in, instead of accessing vast Web directories. The construction of community web directories with usage data collected from the Internet Service Provider raises a number of research issues, which are addressed and solved in this paper. User communities are formed using data collected from the log files of Internet Service Provider. The goal is constructing community web directories based on web usage data. The usage data that form the basis for the construction of community Web directories are collected by the proxy servers of Internet service provider.

## 2. Previous work

A Web directory, such as Microsoft (www.microsoft.com) and the Open Directory Project (ODP) (dmoz.org), users can find any topic that they are interested ,by searching the websites starting with large categories and gradually moving down and choosing the category related to their interests. However, the information for the particular topic ,user has to check deep inside the directory. Hence, the size and the complexity of the Web directory has increased and information overload problem will occur, i.e., it is often difficult to navigate to the information of interest to a particular user. On the other hand, Web Personalization, i.e., the task of making Web-based information systems changed to the needs and interests of individual users, or groups of users, to tackle information overload. However, in achieving personalization, we need to create the accurate and operational user models, but creating and maintaining these models are difficult for various reasons.

## 3. Present Work

In the concept of a community Web directory, a Web directory is created according to the needs and interests of particular user communities. Furthermore, it presents the complete method for the creation of such directories by using web usage data. User community models take the form of thematic hierarchies and are constructed by using clustering approach. We applied our methodology to the ODP directory, and also to an artificial Web directory, which was generated by clustering Web pages that appear in the log file of a proxy server. For the discovery of the community models, we introduced a new criterion that combines a priori thematic informativeness of the Web directory categories with the level of interest observed in the usage data. We have tested the methodology using access logs from the proxy servers of an Internet Service Provider(ISP) and provided results that indicates the usability of the community Web directories. The results presented here provide an initial measure of the benefits that we can obtain by constructing Web directories to the needs and interests of user communities. However, we have only approximated the gain of the end user and have not bothered about the cost of "losses" that could be encountered in the case of personalized web directory. This issue requires the evaluation of community Web directories according to the user preferences.

### 3.1 ISP Log Format

The Common Log Format is a standardized text file format used by web servers when generating access log files. Because the format is standardized, the files may be analyzed by a number of analysis programs. Each line in a file stored in the Common Log Format has the following syntax:

host ident authuser date request status bytes

[edit]

Example

 127.0.0.1 - frank [10/Oct/2000:13:55:36 -0700] "GET /apache_pb.gif HTTP/1.0" 200 2326

A "-" in a field indicates missing data.

127.0.0.1 is the IP address of the client (remote host) which made the request to the server.

- RFC 1413 identity of the client.

frank is the userid of the person who requesting the document.

[10/Oct/2000:13:55:36 -0700] is the date, time, and time zone when the server finished processing the request.

"GET /apache_pb.gif HTTP/1.0" is the request line from the client. The method GET, /apache_pb.gif the resource requested, and HTTP/1.0 the HTTP protocol.

200 is the HTTP status code returned to the client. 2xx indicates successful response, 3xx indicates redirection, 4xx indicates client error, and 5xx indicates server error.

2326 is the size of the object returned to the client, which is measured in bytes.

### 3.1.1 Example of ISP log file

127.0.0.1 - frank [10/Oct/2000:13:55:36 -0700] "GET /apache_pb.gif HTTP/1.0" 200 2326

127.0.0.1 - frank [10/Oct/2000:14:00:36 -0700] "GET /www.microsoft.com/ HTTP/1.0" 200 2326

127.0.0.1 - frank [10/Oct/2000:14:02:36 -0700] "GET /www.microsoft.com/downloads/details.aspx?FamilyId=c22d6a7b-546f-4407-8ef6 d60c8ee221ed&displaylang=en HTTP/1.0" 200 2326

127.0.0.1 - frank [10/Oct/2000:14:03:36 -0700] "GET /www.microsoft.com/contact/contactus.html HTTP/1.0" 200 2326

127.0.0.1 - frank [10/Oct/2000:14:04:36 -0700] "GET /www.google.co.in/search?client=opera&rls=en&q=sample+xml+file&sourceid=opera&ie=utf-8&oe=utf-8&channel=suggest HTTP/1.0" 200 2326

127.0.0.1 - frank [10/Oct/2000:14:04:56 -0700] "GET /www.google.co.in/ HTTP/1.0" 200 2326

127.0.0.1 - frank [10/Oct/2000:14:05:36 -0700] "GET /www.google.co.in/imghp?hl=en&tab=wi HTTP/1.0" 200 2326

127.0.0.1 - frank [10/Oct/2000:14:08:06 -0700] "GET /www.google.com/accounts/ServiceLogin?service=mail&passive=true&rm=false&continue=htt p%3A%2F%2Fmail.google.com%2Fmail%2F%3Fui%3Dhtml%26zy%3Dl&bsv=llya694le36z &scc=1<mpl=default<mplcache=2&from=login HTTP/1.0" 200 2326

127.0.0.1 - frank [10/Oct/2000:14:09:30 -0700] "GET /www.w3schools.com/xml/note.xml HTTP/1.0" 200 2326

127.0.0.1 - frank [10/Oct/2000:14:10:36 -0700] "GET /www.w3schools.com/xml/cd_catalog.xml HTTP/1.0" 200 2326

127.0.0.1 - frank [10/Oct/2000:14:11:26 -0700] "GET /www.w3schools.com/xml/xml_examples.asp HTTP/1.0" 200 2326

127.0.0.1 - frank [10/Oct/2000:14:12:16 -0700] "GET /www.w3schools.com/html/default.asp HTTP/1.0" 200 2326

127.0.0.1 - frank [10/Oct/2000:14:13:36 -0700] "GET /www.w3schools.com/asp/default.asp HTTP/1.0" 200 2326

## 4. Community Web Directory Discovery Algorithm

In this section, we describe the new clustering methodology that we propose for the construction of the community Web directories. This method is used for the selection of categories of the Web directory .

Step1: Load ISP log file.

Step2: Extract information from the log file.

Step3:   Extract websites from log file.

Step4:   Extract directories of the websites.

Step5:   Cluster directories based on training set into various clusters. If cluster is not found then cluster is "unspecified"

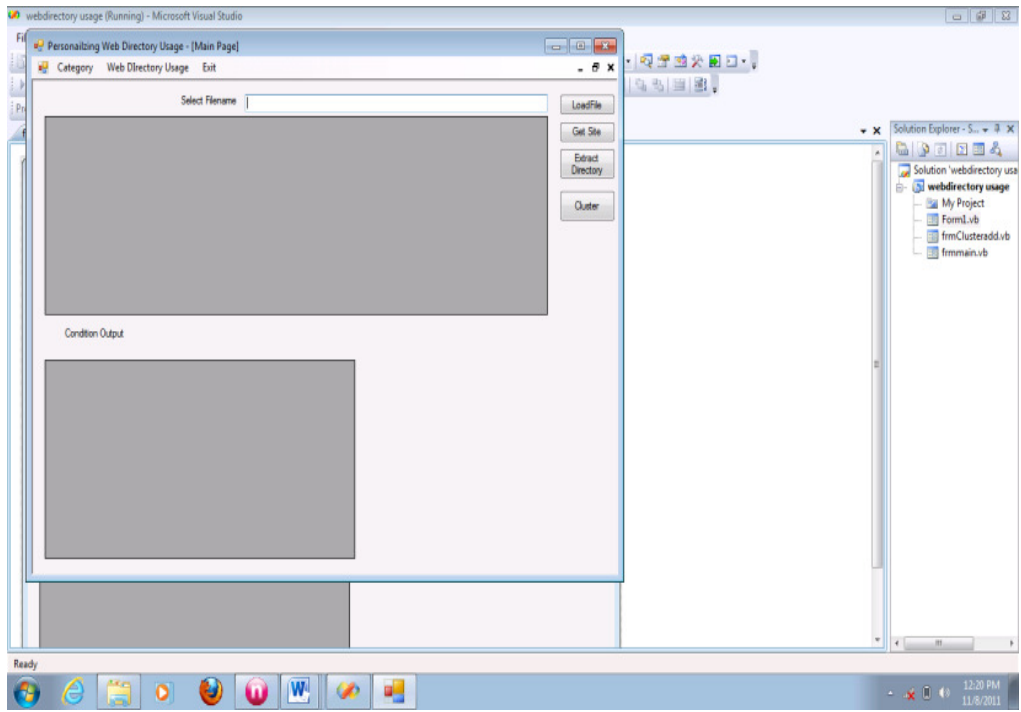

Figure.1 construction of web directory.

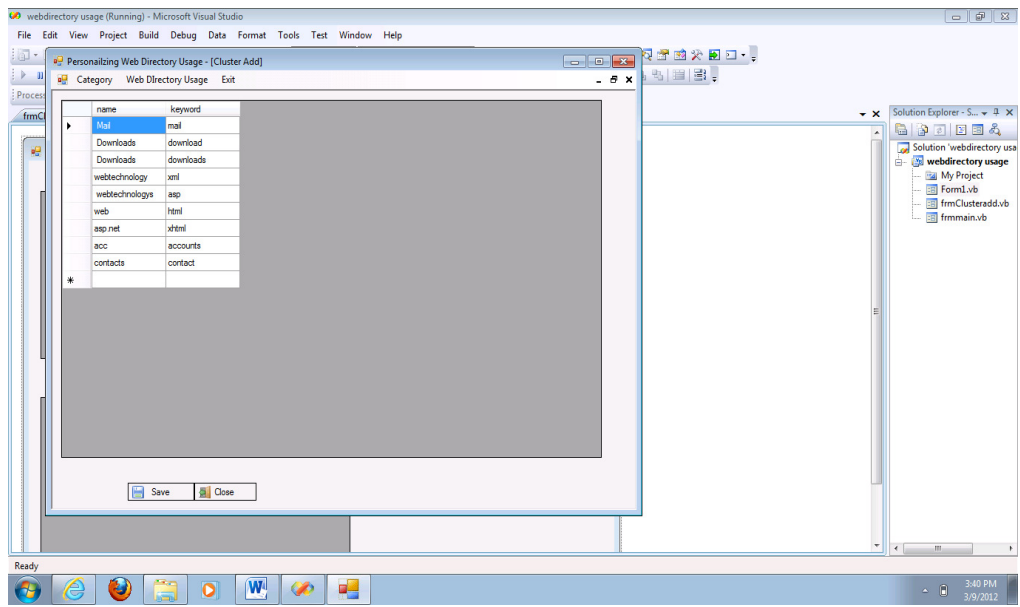

Figure.2 Add and update category.

## 5. Evaluation of Community Web Directories

The methodology introduced in this paper for the construction of community Web directories has been tested for analyzing the data collected from the proxy servers. Evaluation of community web directories involves the following steps.

### 5.1 Experimental Setup

Step1:  Click on load file.

Step2:  Select file i.e ISP log.

Step3:  Click Get Site.

Step4:  Click Extract Directory.

Step5:  Click Cluster.

Step6:  Exit.

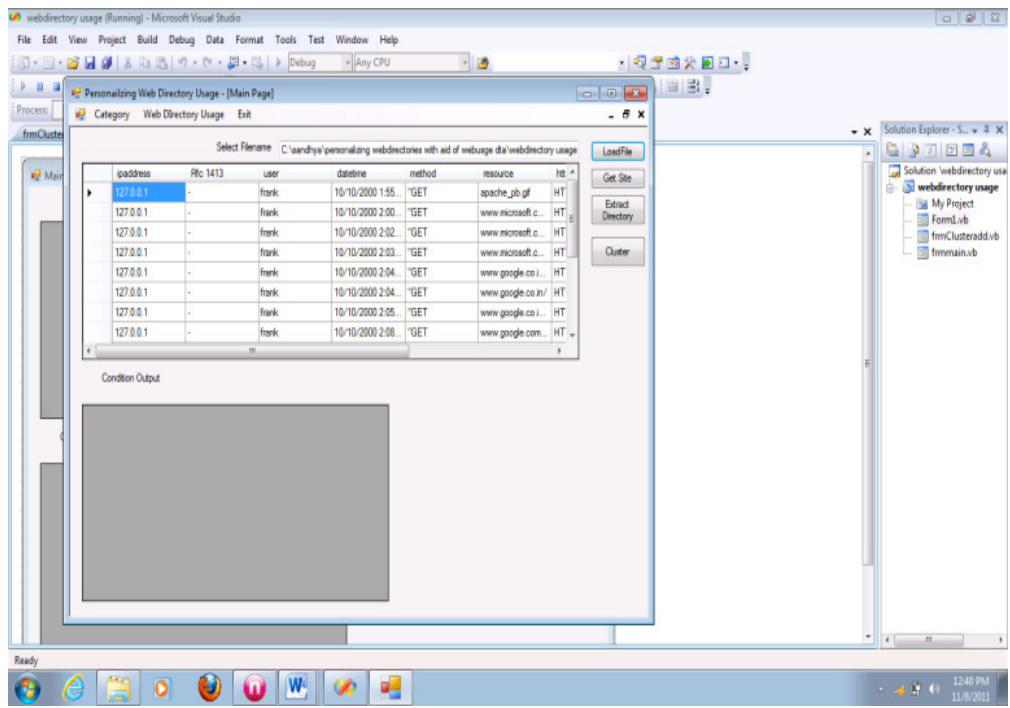

Figure.3  loading log file.

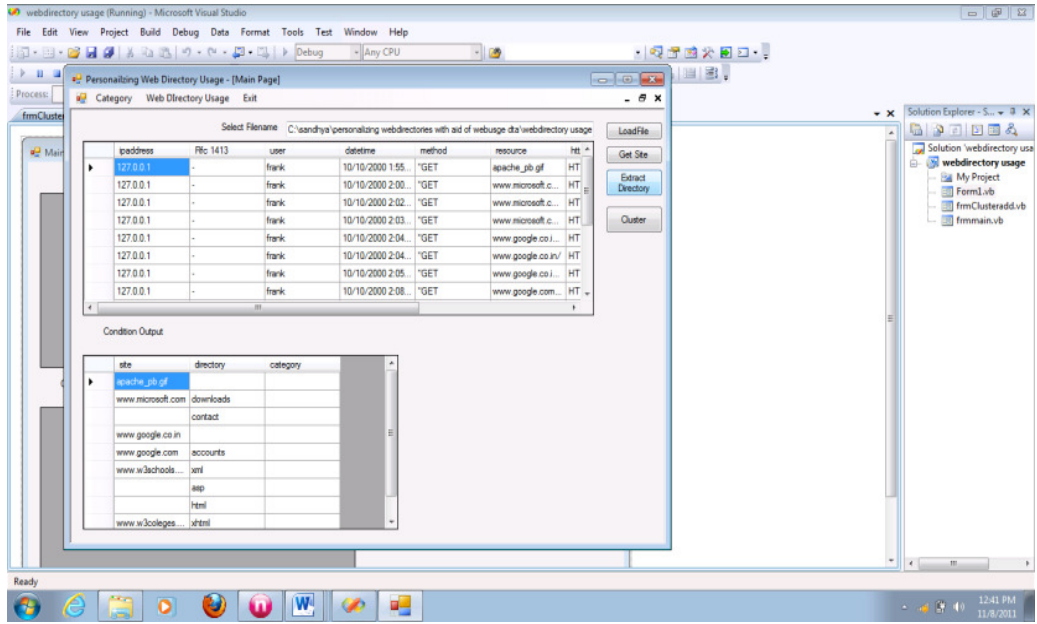

Figure.4 Extracting sites and web directories

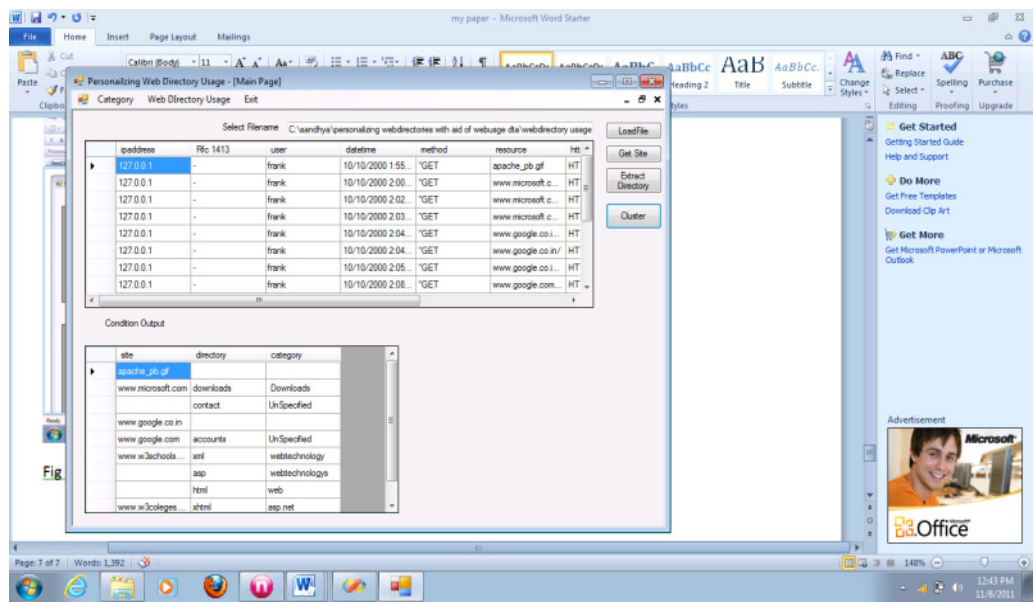

Figure.5 Cluster the directories

## 6. Conclusions and Future Work

We conclude that the concept of community Web directory specializes to the needs and interests of particular user communities. Furthermore, it presents the complete method for the creation of such directories by using web usage data. User community models are constructed by employing clustering approach. We applied our methodology to the ODP directory, as well as to an artificial Web directory, which was generated by clustering Web pages that appear in the access log files of a Web proxy. For the discovery of the community models, we introduced a new criterion that combines a priori thematic informativeness of the Web directory categories with the level of interest observed in the usage data. We have tested the methodology using access logs from the proxy servers of an Internet Service Provider(ISP) and provided results that indicates the usability of the community Web directories. The proposed methodology addresses the issues of existing method by reducing the dimensionality of the problem, through the classification of individual Web pages into the different categories of the web directory. The results presented here provide an initial measure of the benefits that we can obtain by creating Web directories to the needs and interests of user communities. However, we have only approximated the gain of the end user and have not bothered about the cost of "losses" that could be encountered in the case of personalized web directory. This issue requires the evaluation of community Web directories according to user preferences. Various components of the methodology could be replaced by a number of alternatives. Most importantly, more sophisticated methods for extracting the categories from usage data, in addition to the use of an existing Web directory, would make the mapping of pages to domains and then to categories more accurate and complete.

## Acknowledgments

We extremely thank our Principal and the management for their continuous support in Research and Development. We are also very grateful to our faculty members for their valuable suggestions and their ever ending support. Especially, we thank our college Principal and management for their financial support for receiving the sponsorship.

**First Author:** Ramancha Sandhyarani is presently working as Asst. Prof in Computer Science Engineering Department at Jayamukhi Institute of Technological Sciences, Warangal for the past 5 years. She is currently pursuing her Master Technology at Jayamukhi Institute of Technological Sciences, JNTU Hyderabad, (A.P, India).

**Second Author:** Bodakuntla Rajkumar is presently working as Assoc .Professor in Information Technology Department at Jayamukhi Institute of Technological Sciences, Warangal for the past 8 years. He has received his Master of Engineering Degree in System Engineering from Lausitz University (Germany).

**Third Author:** Dr.Jayadev Gyani is presently working as Head and Professor of Computer Science Engineering Department at Jayamukhi Institute of Technological Sciences, Warangal. A.P (INDIA).He has earned his Ph.D. in Computer Science, from Hyderabad Central University; Hyderabad (A.P, India) in 2009.He has 18 years of Teaching Experience in the field of computer science.